\documentclass[superscriptaddress, showpacs, aps, nofootinbib, floatfix, twocolumn]{revtex4-1}
\usepackage[utf8x]{inputenc}
\usepackage{graphicx}
\usepackage[cmex10]{amsmath}
\usepackage{placeins}
\usepackage{color}
\newcommand{\Op}[1]{{\hat{\rm {#1}}}}

\begin{document}

\title{A minimalistic diode}

\author{Tim Collet}
 \affiliation{Institute for Theoretical Condensed Matter physics, Karlsruhe Institute of Technology, 76128 Karlsruhe, Germany}
 \email{tcollet@tkm.uni-karlsruhe.de}
\author{Peter Schmitteckert}
 \affiliation{Institute of Nanotechnology, Hermann-von-Helmholtz-Platz 1, 76344 Eggenstein-Leopoldshafen, Germany}
 \email{peter.schmitteckert@kit.edu}
 \pacs{PACS-Nummer}
\date{\today}

\begin{abstract}
In this paper we present a minimalistic model sufficient for current asymmetries in molecules.
In particular, we search for an interaction on the molecule which causes current asymmetries independent of additional assumptions or the precise form of its environment.
To this end we first discuss earlier proposals and clarify the importance of additional assumptions.
We then present a minimal model of a strongly polarizable molecule which shows a strongly asymmetric I/V calculated within time dependent DMRG.
\end{abstract}

\maketitle

\section{Introduction}

Diodes are a base ingredient to electronics,
so the development of new molecule based diodes could improve molecular electronics.
The first theoretical proposal of a molecule based diode was given in \cite{aviram} by Aviram and Ratner (AR),
which to our knowledge has not been realized experimentally yet.
Since then, there have been many experimental realizations of molecular diodes \cite{exreali1, exreali3, exreali4, exreali5, exreali6, exreali7},
which have mainly used strongly polarizable molecules.
However, the mechanisms causing rectification are still being discussed \cite{discussion1, 2002PhRvB..66p5436K, doi:10.1021/ja028229x, Datta}.
One ansatz to investigate these mechanisms is to embed particular molecular properties into simple models,
and to investigate their ability to cause current asymmetries.
To obtain reliable diodes, these asymmetries should be independent of the specific form of the environment of the molecule.\\ 
In the proposal by AR, interactions were implicitly assumed but not written down explicitly in the Hamiltonian.
This does not allow to single out a specific interaction causing current asymmetries.
Also, there are additional assumptions on the form of the environment of the molecule needed.\\
Another theoretical proposal we want to investigate was given in \cite{roy}.
There, all assumed interactions are explicitly written down in the Hamiltonian.
However, the author had to assume restrictions on the form of the band structure of the leads appended to the molecule in order to find current asymmetries.\\
In this paper, we present a model of a molecular diode where the interaction needed is specified in the Hamiltonian of the molecule,
and the current asymmetry is observed independent of the exact specifications of the leads.\\
In section \ref{methods} we start with a description of the methods used to calculate the current.
In section \ref{AR} we then review the proposal by
AR
in the context of our topic.
Section \ref{Roy} deals with the other proposal mentioned above 
and validates the statement by D. Roy that a current asymmetry is observed only for certain conditions in the leads in this model.
Our model, as well as results for the current in this model, are presented in section \ref{own}
before we conclude in section \ref{conclusion}.

\section{Methods}
\label{methods}
We define the quantity 
\begin{equation}
 A \left(V \right) = \frac{I \left(V \right) + I \left(-V \right)}{I \left(V \right) - I \left(-V \right)},
\end{equation}
with $I$ being a current, as the current asymmetry.
To measure a current in the observed systems, we apply two physically different quenching protocols 
used 
for example in \cite{ANDP:ANDP201000017}:
In both cases we quench at zero time in the chemical potentials of the leads applied to the molecule.
The quench then effectively adds to or removes from the Hamiltonian the charge imbalance, or voltage, term
\begin{equation}
 \mathcal{H}_V = V/2 \left( \sum_{x}^{ \text{left lead}} \Op n_x - \sum_{x}^{ \text{right lead}} \Op n_x \right)
\end{equation}
with $V$ being the applied bias voltage and $\Op n_x$ being the particle density at site x\footnote{Since we do not investigate spin interactions, we freeze out the spin of the fermions in our calculations, effectively looking at spinless fermions.}.
If we start with leads at different chemical potentials and quench the potentials to be equal for times larger than zero,
the voltage influences the initial density distribution.
This situation corresponds to  scattering theory calculations where particles are injected in one lead
and will either be reflected in the same lead or transmitted to the other lead.
In those calculations, the effect of the energy dispersion gets cancelled by the density of states.
In contrast, if $ \mathcal{H}_V$ is turned on 
at $t = 0$, the resulting current will
be influenced by the finite band width. E.g.\ the current vanishes for voltages larger than the band width, for details see \cite{ANDP:ANDP201000017}.

Our goal will be to find current asymmetries using the former method, since those asymmetries will not depend on the band structure of the leads.\\
Within this paper we can use the different quenching schemes to judge sensitivity to details of the lead structure.

The initial state of the system is obtained as the ground state of the system before the quench via finite lattice DMRG \cite{PhysRevLett.69.2863}.
The time evolution of this state after the quench is calculated using the time dependent DMRG (td-DMRG) and the Krylov space approximation
for the matrix exponential as described in \cite{PhysRevB.70.121302,PhysRevLett.101.140601,ANDP:ANDP201000017}.

\begin{figure}[t]
 \includegraphics[width=0.45\textwidth]{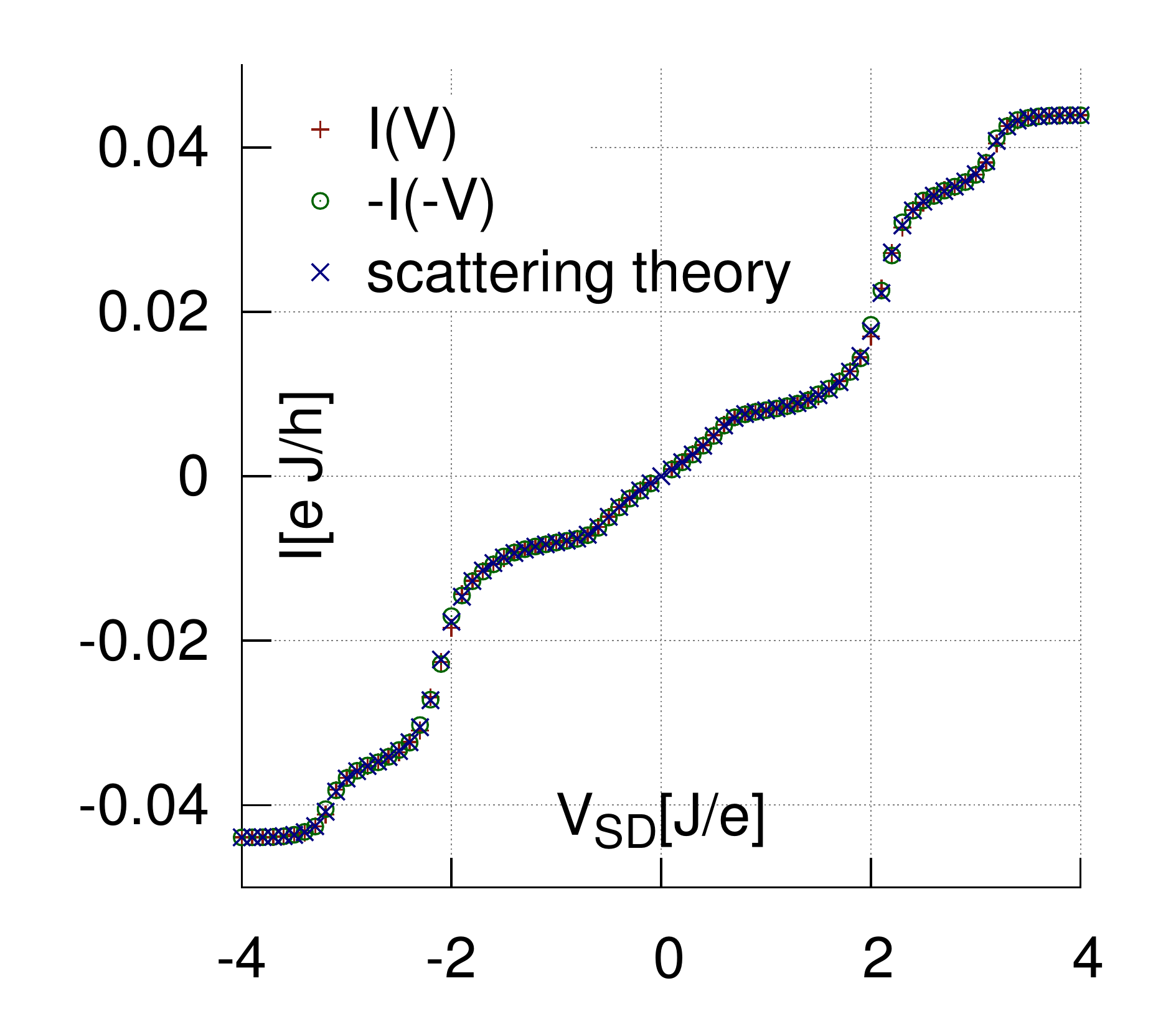}
 \caption{Result for the model \eqref{ARham}, calculated via scattering theory (the blue 'x' es) and via exact diagonalization (the red '+' es).
 The green circles denoting $-I(-V)$ are meant to visualize the size of the asymmetry.
 Parameters are $J_{1} = 0.3 J$, $J_{2} = 0.09 J$, $\mathrm{w}_{0} = -J$, 
        $\mathrm{w}_{1} = 0.3 J$, 
        $\mathrm{w}_{2} = - 0.3 J$, 
        $\mathrm{w}_{3} = 1.5 J$ and for the diagonalization 
        $M = 500$.}
\label{AR-Res}
\end{figure}
\begin{figure}
 \includegraphics[width=0.45\textwidth]{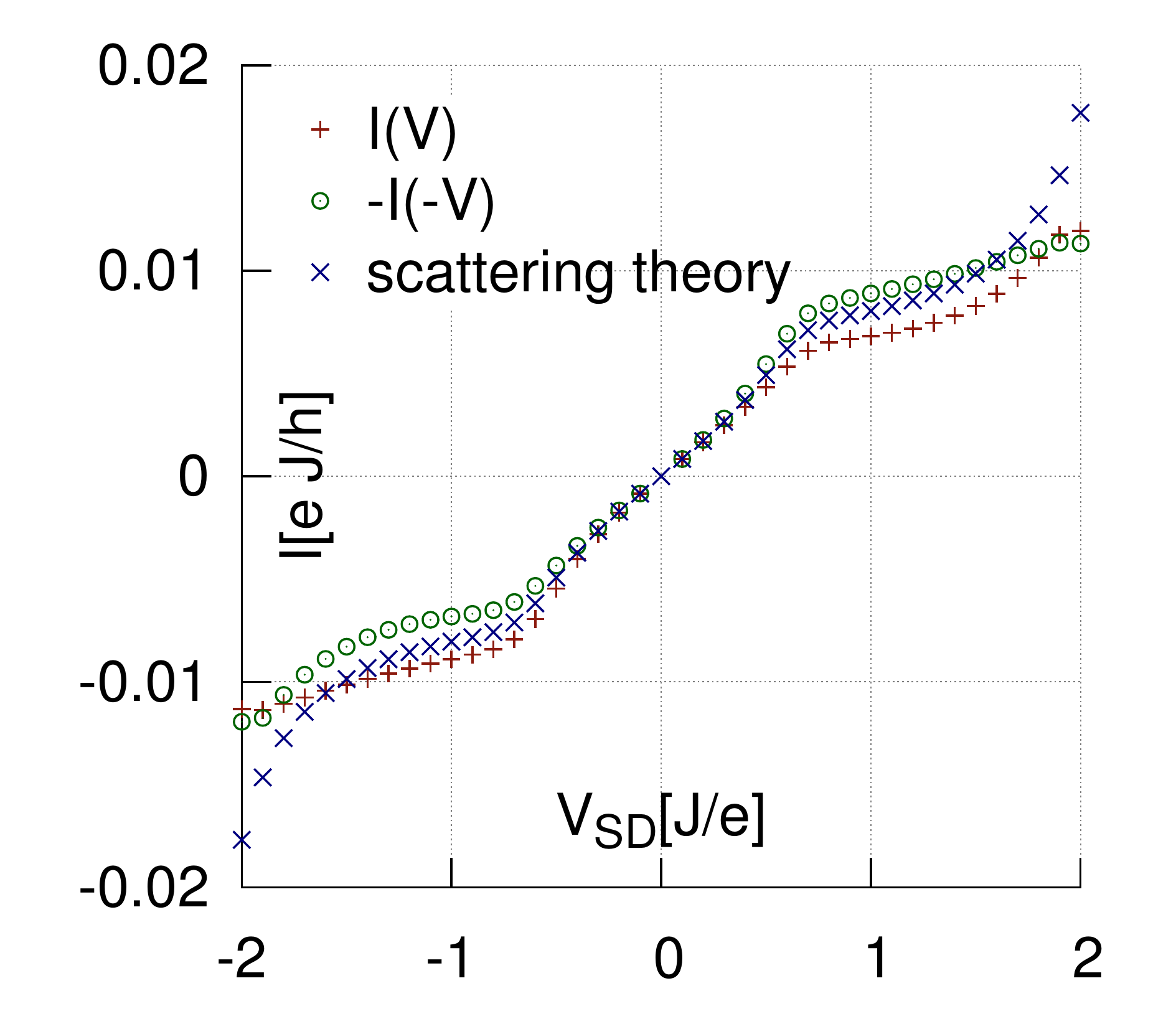}
 \caption{Result for the model \eqref{ARham} when quenching to different chemical potentials in the leads. 
 The blue 'x' es are the analytic results from FIG. \ref{AR-Res}.
 Parameters are $J_{1} = 0.3 J$, $J_{2} = 0.09 J$, $\mathrm{w}_{0} = -J$, 
        $\mathrm{w}_{1} = 0.3 J$, 
        $\mathrm{w}_{2} = - 0.3 J$,
        $\mathrm{w}_{3} = 1.5 J$,
        timesteps $\Delta t = 0.25 J$ and
        System size $M = 1000$.}
\label{AR-Res2}
\end{figure}

The resulting current expectation value reaches a plateau with some small oscillations after some transient regime (initial adjusting), and later drops due to reflection of the charge density wavefront.
The value of the plateau gives us the current in the large lead limit, for details see \cite{ANDP:ANDP201000017}.
In the case of the non-interacting model in section \ref{AR}, we use a free fermion picture for scattering theory calculations as well as exact diagonalization instead of the DMRG procedure.
Since we can only compute finite systems in numerical calculations, we will set our system size to $M$ instead of $\infty$.
This means that e.g. 
for systems with probes restricted to one site, the summation boundaries of the leads 
are $- \frac{M}{2}$ instead of $- \infty$ and $\frac{M}{2} - 1$ instead of $\infty$.
In cases where the probe consists of an odd number of sites, the size of the leads differ in our calculations by one site 
since we have to use an even number of sites to obtain an integer particle number for half or quarter filling.
An odd number of those sites models our probe, so we are left with an odd number of sites for two leads.
For finite size estimates we 
compare the current obtained in a system with swapped leads to the original current.

\section{The Aviram-Ratner proposal}
\label{AR}

In their paper,
AR
looked at a class of molecules 
effectively 
consisting of two two level systems separated by a bridge part.
\begin{figure}[h]
 \includegraphics[width=0.45\textwidth]{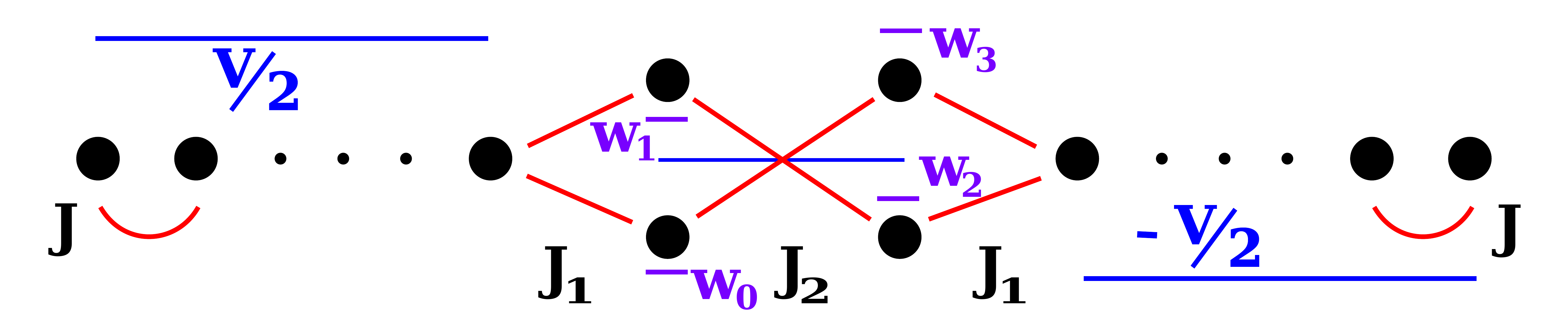}
 \caption{Sketch of the system modelled by \eqref{ARham}. Red lines connecting dots describe hopping terms, blue, long lines describe the quenched chemical potential, and purple, short lines the energy levels of the two level systems. }
\label{ARsketch}
\end{figure}
One of the two level systems is an acceptor, with the upper level having slightly higher energy than the Fermi surface.
The other one is a donor, with the lower level being closely below the Fermi surface.
The authors implicitly assumed a screening interaction which causes the chemical potential to increase or decrease linearly between the two leads.
They also assumed that electrons would relax after hopping onto a two level system.
Now they argued that for shifted chemical potentials, significant current would start to flow from acceptor to donor 
once the donor lower level was higher\footnote{Energy measured relative to the chemical potential.} than the Fermi surface in the nearest lead,
while a significant current flow in the other direction would require the donor lower level to  be at least of the same height as the acceptor upper level
\footnote{The requirement could also be that the upper level of the donor is lower than the Fermi surface in the lead, depending on parameters.}.
As argued e.g. in chapter 5 of \cite{0957-4484-15-7-051}, the influence of screening depends on the form of the leads.
Since we want to find asymmetries independent of external factors, we neglect screening interactions. 
The same goes for 
relaxation effects.

Also, \cite{Datta} suggests that the assumption of a linear change in chemical potential between the leads might be wrong.
In the remaining part of the section, we show that in scattering theory without the assumption of additional screening and relaxation
the AR system does not rectify currents.\\
The Hamiltonian we assume for the system with constant chemical potential is sketched in FIG. \ref{ARsketch} and reads
\begin{align}
        \mathcal{H} = & -J \sum_{x=-\infty}^{-1} \left( \mathrm{c}^{\dagger}_{x}\mathrm{c}_{x-1}  \right) - J \sum_{x=5}^{\infty} \left( \mathrm{c}^{\dagger}_{x}\mathrm{c}_{x-1}  \right) \nonumber \\
        & - J_{1} \left( \mathrm{c}^{\dagger}_{0}\mathrm{c}_{-1} + \mathrm{c}^{\dagger}_{1}\mathrm{c}_{-1} + \mathrm{c}^{\dagger}_{4}\mathrm{c}_{3} + \mathrm{c}^{\dagger}_{4}\mathrm{c}_{2} \right) \nonumber \\
        & - J_{2} \left( \mathrm{c}^{\dagger}_{2}\mathrm{c}_{0} + \mathrm{c}^{\dagger}_{3}\mathrm{c}_{0} + \mathrm{c}^{\dagger}_{2}\mathrm{c}_{1} + \mathrm{c}^{\dagger}_{3}\mathrm{c}_{1} \right) + \text{h.c.}  \nonumber \\
        & + \sum_{\mathrm{i}=0}^{3} \mathrm{w}_{\mathrm{i}} \hat{\mathrm{n}}_{\mathrm{i}} \label{ARham}.
\end{align}

\begin{figure}[t]
 \includegraphics[width=.45\textwidth]{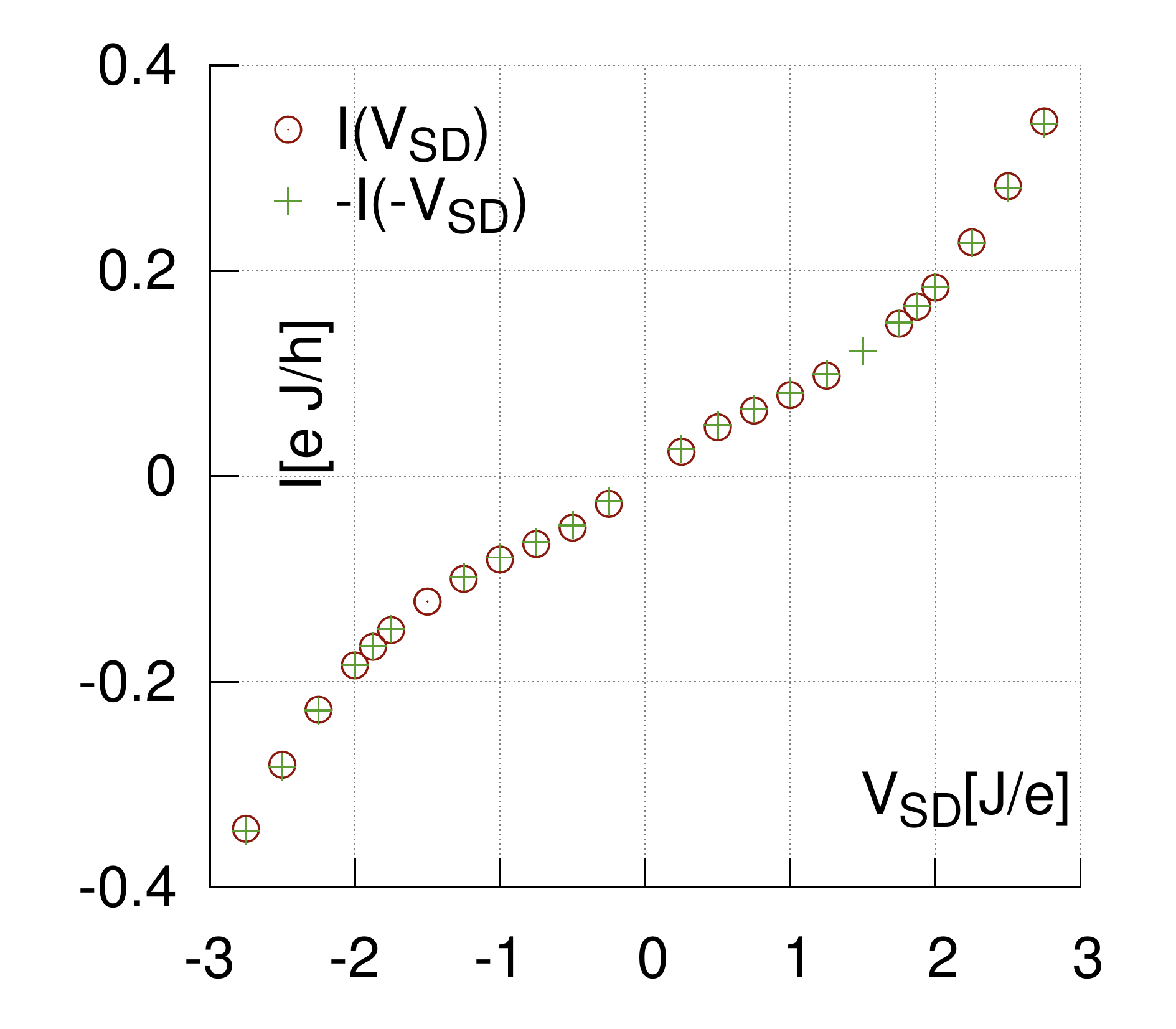}
 \caption{Current vs.\ bias voltage in an IRLM system for a total system size $M = 140$, quarter filling, left lead to impurity density density interaction $J_{nn_1} = 0.2 J$, system to right lead density densityinteraction $J_{nn_2} = 0.5 J$, 
  timesteps $\Delta t = 0.25 \hbar/J$ and system to lead hopping $J_{c} = 0.5 J$.
  Here, bias voltage affected only the time evolution Hamiltonian.}
\label{asyInterHtime}
\end{figure}

\begin{figure}[t]
 \includegraphics[width=.45\textwidth]{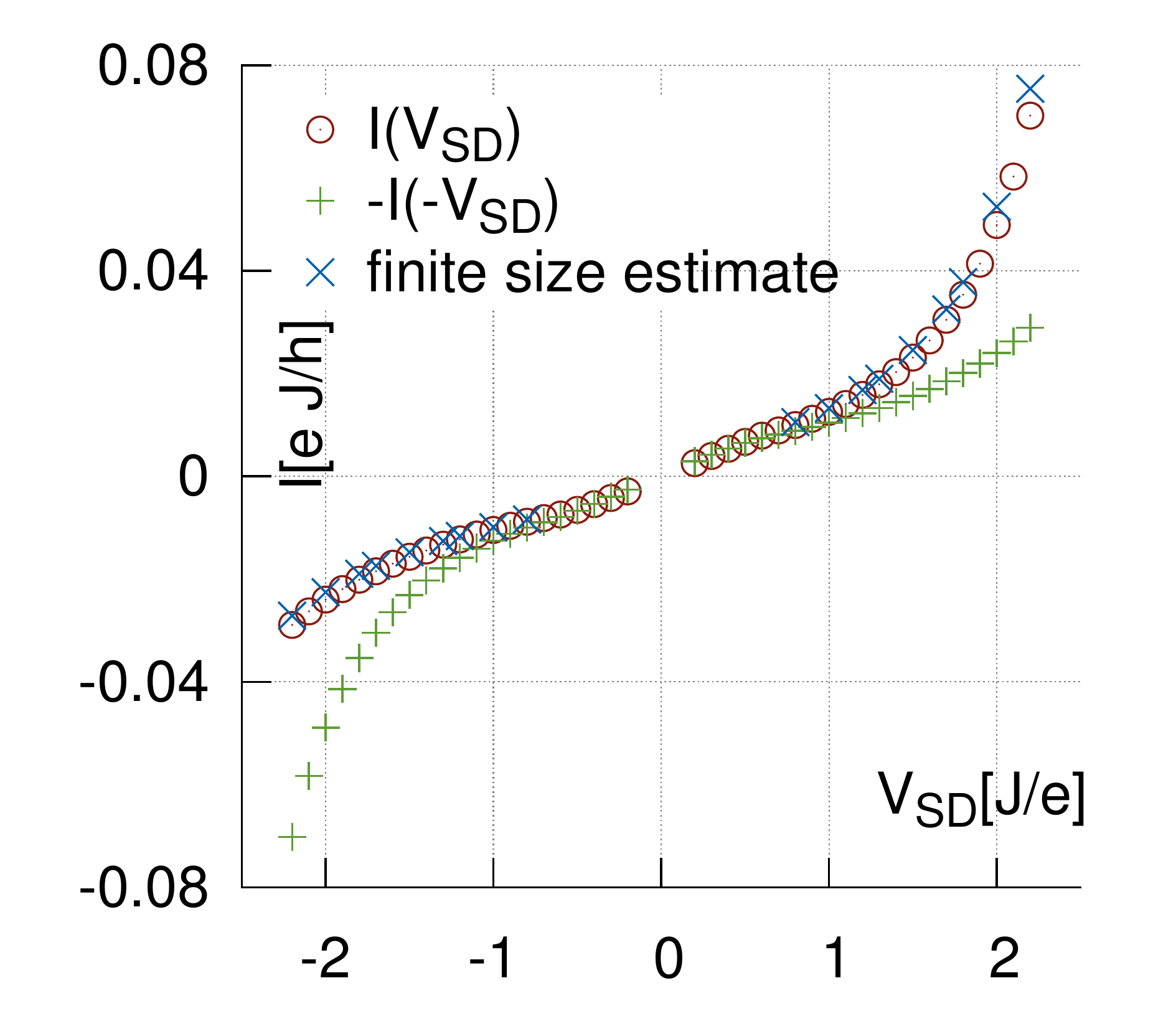}
 \caption{Current vs.\ bias voltage in an IRLM system of total system size $M = 140$, quarter filling, density density interaction $J_{nn_{1/2}} = 0.5J$, time steps $\Delta t = 0.25 \hbar/J$, left lead to system hopping $J_{c_1} = 0.2J$ and system to right lead hopping $J_{c_2} = 0.5J$.
 Bias voltage is included in the time evolution Hamiltonian. 
 The red 'o' es mark the current itself, the green '+' es mark -I(-U) to visualize the size of the asymmetry, and the blue 'x' es mark the current for swapped left and right lead to give an estimate of the finite size effects.
}
\label{asyJuncHtime}
\end{figure}

Here, $\mathrm{c}^{\dagger}_{x}$ and $\mathrm{c}_{x}$ are creation and annihilation operators at site x, $J$ is the lead hopping,
$J_1$ denotes hopping from leads to the two level systems, $J_2$ hopping between the two level systems and the $\mathrm{w}_{\mathrm{i}}$ 
are the energies of the different energy levels.

Making a plane wave ansatz for the eigenfunctions
\begin{align}
         \mathrm{c}^{\dagger}_{k} & = \sum_{x=-\infty}^{-1} \left( \mathrm{e}^{i k x} + \mathrm{r} \mathrm{e}^{-i k x} \right) \mathrm{c}^{\dagger}_{x} \nonumber \\
                                  & + \sum_{x=4}^{\infty} \mathrm{t} \mathrm{e}^{i k x} \mathrm{c}^{\dagger}_{x} \nonumber \\
                                  & + \sum_{\mathrm{i}=0}^{3} a_{i} \, \mathrm{c}^{\dagger}_{\mathrm{i}} ,
\end{align}
$r$ being the reflection and $t$ the transmission amplitude, and $a_{i}$ are the amplitudes on the two level systems,
we solve for the transmission amplitude with the requirement of the ansatz being an eigenstate with eigenenergy $E_k$:
\begin{equation}
  \left[ \mathcal{H}, \mathrm{c}^{\dagger}_{k} \right]  = E_k \mathrm{c}^{\dagger}_{k} .
\end{equation}
From there we calculate the current via \cite{Landauer1,doi:10.1080/14786437008238472,PhysRevLett.57.1761}
\begin{equation}
  \mathrm{I} = \frac{e}{{h}} \int_{ - \frac{V}{2} }^{ \frac{V}{2} } {\mathrm d} E \left| t \left( E \right) \right|^2.
\end{equation}

The results obtained from this scattering theory ansatz and exact diagonalization are shown for one set of parameters in FIG. \ref{AR-Res}.
We use exact diagonalization for a finite system size $M$ to calculate the current for the quench to different chemical potentials.
The result for the according simulation with $\mathcal{H}_V $ in the time evolution operator is shown in FIG. \ref{AR-Res2}.
For chemical potentials outside the range of $V \in  \left[-2 J, 2 J\right]$, we could not obtain reliable results due to a very slowly decaying transient current.
While one sees a slight asymmetry the latter case for large voltages, there is no asymmetry in the scattering theory calculation 
or in the latter case for small voltage. These results demonstrate that that the system described by the Hamiltonian of Eq.~\ref{ARham}
is not sufficient to describe a diode, the additional assumptions made by AR are essential.
\begin{figure}[t]
 \includegraphics[width=.45\textwidth]{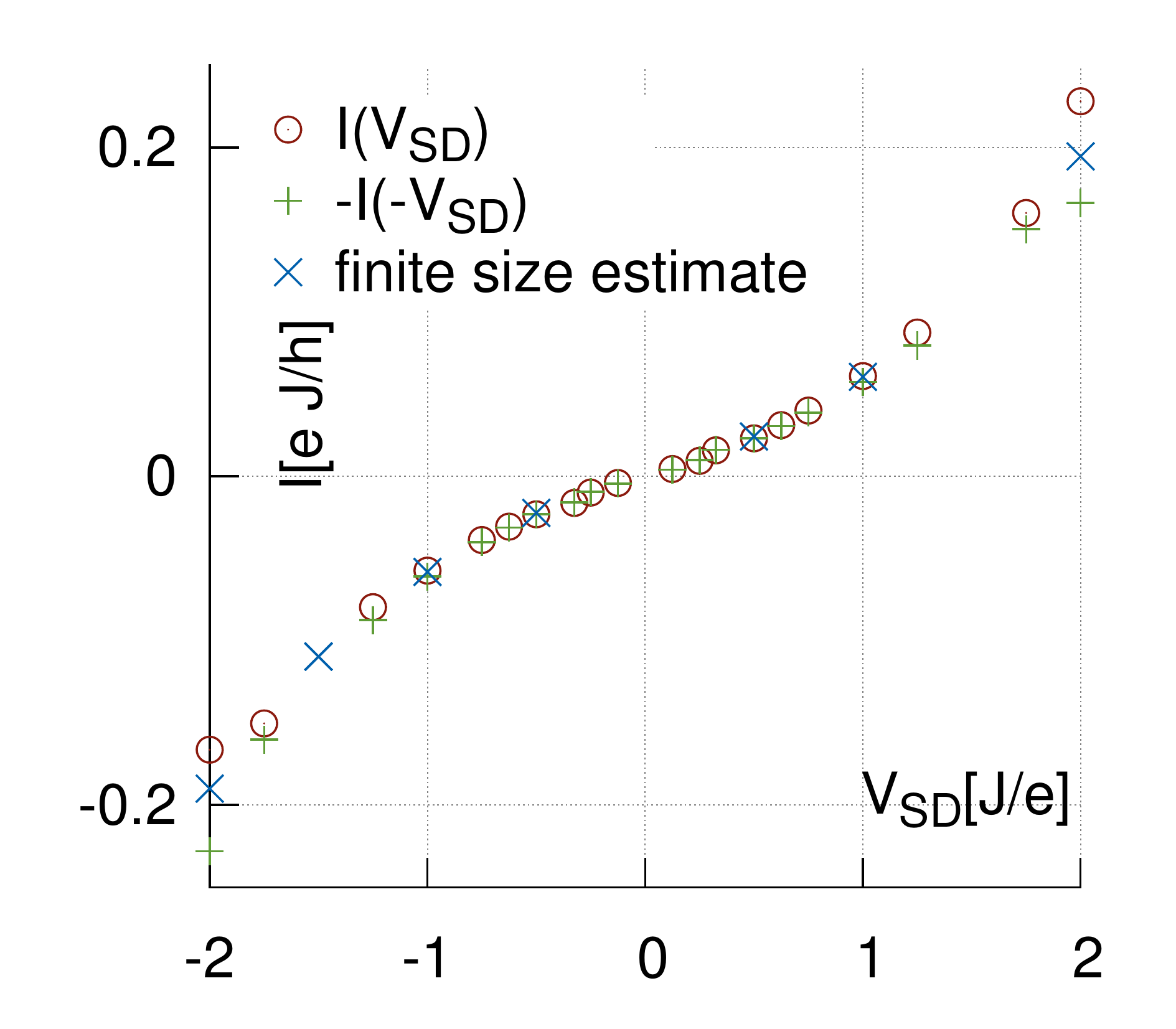}
 \caption{Current depending on bias voltage in IRLM system for system size $M = 130$, quarter filling, left lead to system hopping $J_{c_1} = 0.2$, system to right lead hopping $J_{c_2} = 0.8$, timesteps $\Delta t = 0.25 J$ and density interaction strength $J_{nn_{1/2}} = 0.5 J$.
The red 'o' es mark the current itself, the green '+' es mark -I(-U) to visualize the size of the asymmetry, and the blue 'x' es mark the current for swapped left and right lead to give an estimate of the finite size effects. 
In this calculation the bias voltage entered the Hamiltonian for the initial state. }
\label{otherproposal}
\end{figure}

\section{The Roy model}
\label{Roy}
\begin{figure}
 \includegraphics[width=.45\textwidth]{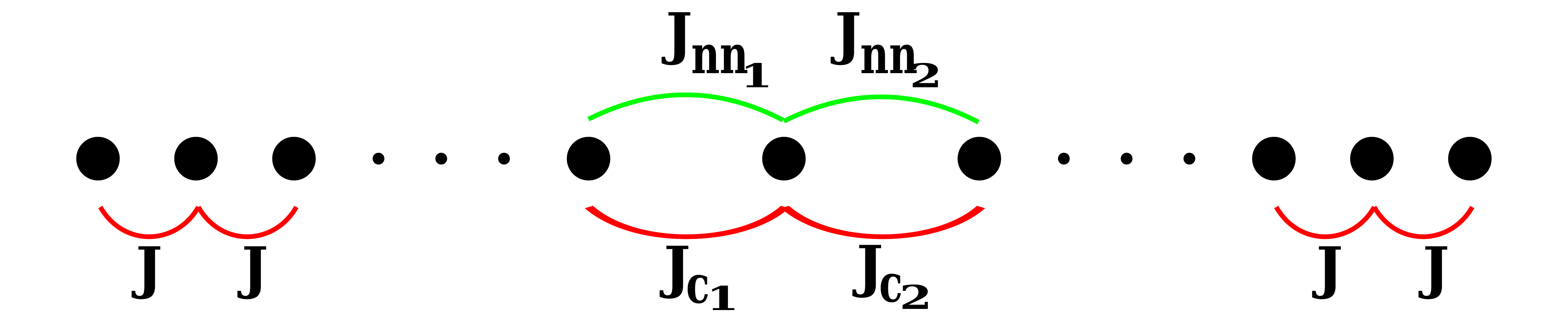}
 \caption{Sketch of the model used in section \ref{Roy}. Red lines below the dots denote hopping, the green lines above density density interaction.}
\label{roysketch}
\end{figure}

The Hamiltonian of the model of Roy \cite{roy} corresponds to an interacting resonant level model (IRLM) with asymmetric couplings, 
which is also sketched in FIG. \ref{roysketch}, and reads:
\begin{align}
  \mathcal{H} = & -J \sum_{x=-\infty}^{-2} \left( \Op{c}^{\dagger}_{x}\Op{c}_{x+1} + \Op{c}^{\dagger}_{x+1}\Op{c}_{x}  \right) \nonumber \\
 & - J \sum_{x=1}^{\infty} \left( \Op{c}^{\dagger}_{x}\Op{c}_{x+1} + \Op{c}^{\dagger}_{x+1}\Op{c}_{x}  \right) \nonumber \\
 & - J_{c_2} \left( \Op{c}^{\dagger}_{0}\Op{c}_{1} + \Op{c}^{\dagger}_{1}\Op{c}_{0}  \right)  - J_{c_1} \left( \Op{c}^{\dagger}_{-1}\Op{c}_{0} + \Op{c}^{\dagger}_{0}\Op{c}_{-1}  \right) \nonumber \\
 & - J_{nn_1} \left( \Op n_{-1} \, \Op n_{0} \right) - J_{nn_2} \left( \Op{n}_{0}\, \Op{n}_{1} \right) \,,
\end{align}

with $\Op n_{x}$ being the particle number operator at position $x$ and the different $J$'s being various hopping and coupling parameters.
The prediction is that asymmetric junctions, which means $J_{c_1} \neq J_{c_2}$, 
and a nonlinear dispersion relation would lead to an asymmetric I-V-curve.
We increased the importance of nonlinear terms in the cosine dispersion relation of the tight binding leads 
by filling the leads only up to crystal momentum $k = \frac{\pi}{4}$ instead of $\frac{\pi}{2}$. \\
The results of our calculations are displayed in in FIGS.~\ref{asyInterHtime}-\ref{otherproposal}.
In FIG.~\ref{asyInterHtime} we compare  the $I(V)$  and $I(-V)$ characteristic obtained by time dependent simulations, where the charge imbalance
is put into the time evolution operator. As predicted by Roy we do not find an asymmetry in the $I/V$ characteristic.
This also holds true for the case of voltage influencing the initial density distribution.
In FIG.~\ref{asyJuncHtime} we show that, consistent with Roy's prediction, there is an asymmetry in the case of symmetric interactions and asymmetric junctions,
provided the charge imbalance term is put into the time evolution operator.
If the charge imbalance determines the initial density distribution and therefore the band structure does not influence the result, this asymmetry reduces to the order of the finite size effects.
This is shown in FIG. \ref{otherproposal}.
As a consequence, we infer that the asymmetry depends heavily on external factors.



\section{Proposed model and Results}
\label{own}
\begin{figure}
 \includegraphics[width=.45\textwidth]{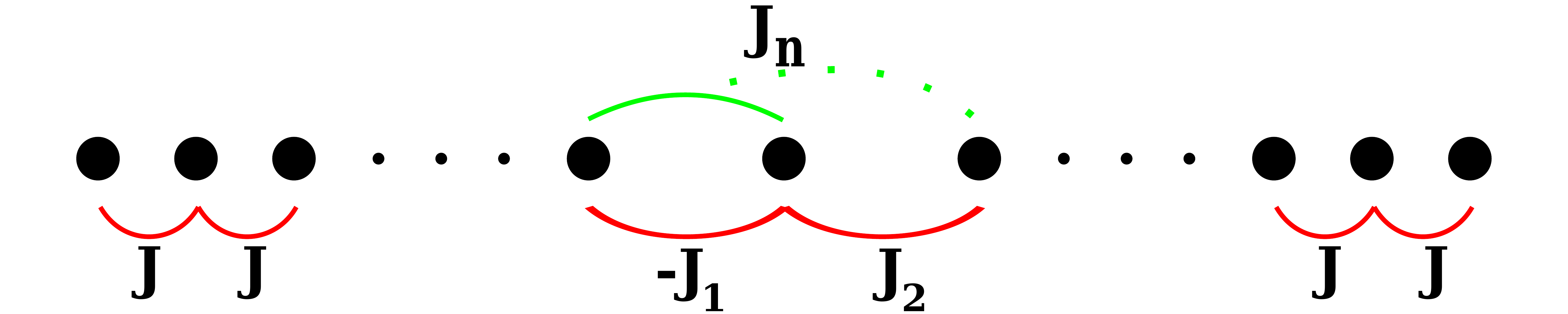}
 \caption{Sketch of the model used in section \ref{own}. Red lines below the dots denote hopping, the green lines above the correlated hopping term.}
\label{ownsketch}
\end{figure}

In order to find a model which causes an asymmetric $I/V$ under scattering theory, independent of specific lead properties, 
it is instructive to look at the way how scattering theory works.
In scattering theory \cite{Lippmann_Schwinger:PR50} one searches for eigenstates of the Hamiltonian which are asymptotically given by incoming
and outgoing (typically plane) waves. For transport calculations this corresponds to occupied eigenstates  of one lead, and empty eigenstates of the other
lead before being connected to the structure under investigation.
The voltage enters via the different electrochemical potentials of the two leads.
Within the Landauer description \cite{Landauer1,doi:10.1080/14786437008238472,PhysRevLett.57.1761} the current is given by
\begin{equation}
  I = \int_{\mu_{\mathrm R}}^{\mu_{\mathrm L}}  T(E) {\rm d}E \,
\end{equation}
where $T(E)$ is the transmission at energy $E$ and the voltage is given by $\mu_{\mathrm L} - \mu_{\mathrm R} $. 
As long as the transmission does not depend on the voltage, the current only depends on it via the integration borders, so scattering theory gives an antisymmetric I/V,
at least for systems with time reversal symmetry.
In order to obtain a symmetric contribution to the $I/V$ characteristic we introduce a strong polarizability via a correlated hopping interaction.
We embed it into a noninteracting resonant level model,
obtaining the Hamiltonian of the system sketched in FIG. \ref{ownsketch}:
\begin{align}
  \mathcal{H} = & -J \sum_{x=-\frac{M}{2}}^{-2} \left( \Op{c}^{\dagger}_{x}\Op{c}_{x+1} + \Op{c}^{\dagger}_{x+1}\Op{c}_{x}  \right) \nonumber \\
 & - J \sum_{x=1}^{\frac{M}{2}} \left( \Op{c}^{\dagger}_{x}\Op{c}_{x+1} + \Op{c}^{\dagger}_{x+1}\Op{c}_{x}  \right) \nonumber \\
 & - J_{2} \left( \Op{c}^{\dagger}_{0}\Op{c}_{1} + \Op{c}^{\dagger}_{1}\Op{c}_{0}  \right) \nonumber \\
 & - \left( J_{n} \, \Op n_{1} - J_{1} \right) \left( \Op{c}^{\dagger}_{-1}\Op{c}_{0} + \Op{c}^{\dagger}_{0}\Op{c}_{-1}  \right) .
\end{align}
\begin{figure}[t]
 \includegraphics[width=.45\textwidth]{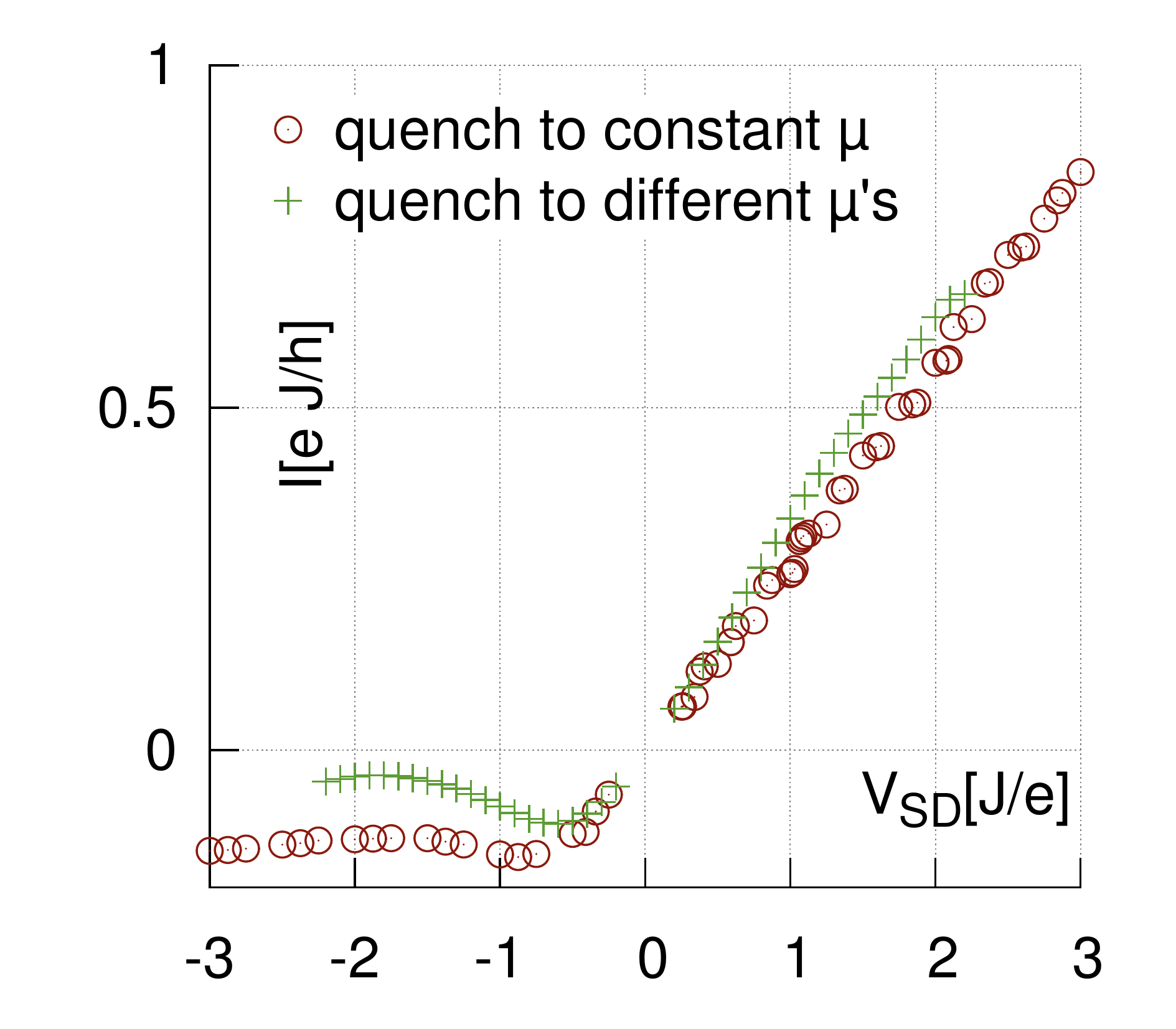}
 \caption{Result for the value of the current plateau (measured at $x = 0$) for $J_{1} = 0.11 \cdot J$, $J_{2} = 0.5 \cdot J$, $J_{n} = 0.5 \cdot J$, system size $M = 110$ and timestep size $\Delta t = 0.25 \cdot J$. }
\label{diode}
\end{figure}

\begin{figure}
 \includegraphics[width=.45\textwidth]{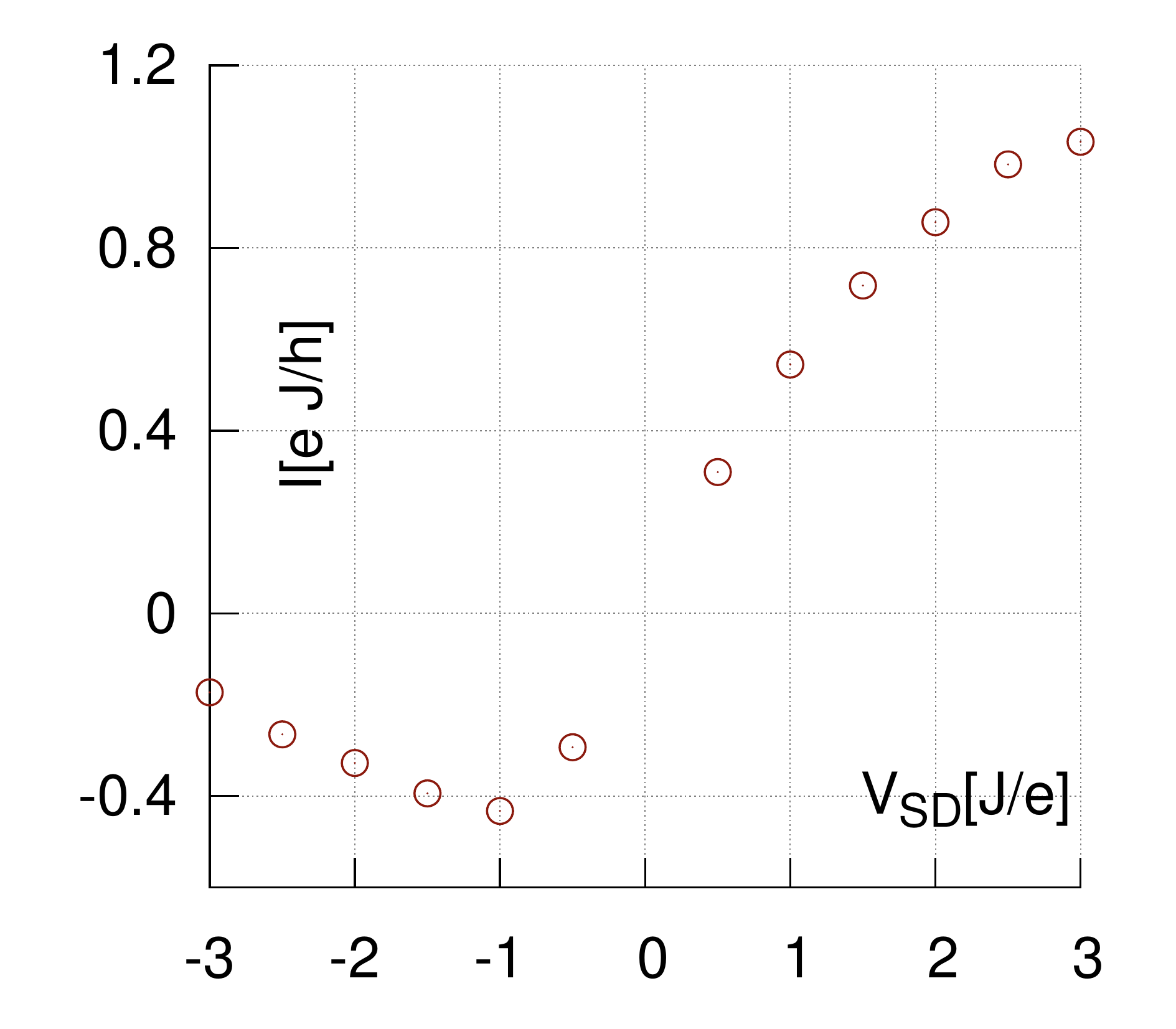}
 \caption{Current depending on bias voltage for $J_{1} = 0$, $J_{2} = 0.5 \cdot J$, $J_{n} = -0.5 \cdot J$, system size $M = 50$ and time step size $\Delta t = 0.25 \cdot J$. }
\label{pure_interaction}
\end{figure}

Numerical computation of the current for different voltages gives the result shown in FIG. \ref{diode}.
The Steps in the I-V-curve are finite size effects\footnote{These effects are due to the finite size of the leads, 
resulting in finite energy levels.} and get smoother for bigger systems.
The parameters in this curve are tuned in a way to maximize similarity to a diode.
In general we find a mix of the standard resonant level model result - approximately an arcus tangens curve - 
and an antisymmetric function\footnote{See fig. \ref{pure_interaction}.} with minimum at around $V_{SD} = - J$, with weights depending on the choice of $J_{1}$ and $J_n$.
We want to point out that the asymmetry in FIG.\ref{diode} is visible for both kinds of quenches applied.
 
\section{Conclusion}
\label{conclusion}
Our goal in this paper was to find hints at interactions or other mechanisms which reliably cause a diode like behaviour independent of external factors.
Revisiting the proposals in \cite{aviram} and \cite{roy}, 
the asymmetries we found in the I/V were crucially dependent on the band structure of the leads. 
However for devices, one desires systems where the properties are given by the system itself and are not being crucially dependent on the structure of the leads.
Here we presented a minimalistic structure that resembles a Diode independent on the quenching schemes. From this we conclude that
its $I/V$ characteristic is given by the structure itself and that it is robust against band structure properties of the leads.
We achieved this property by introducing a correlated hopping interaction leading to strong polarizability of the system.

\section*{Acknowledgements}
We would like to thank Ferdinand Evers for insightful discussions. TC thanks Guy Cohen for discussions about rectifying models.
Parts of this work was performed on the computational resource bwUniCluster funded by the Ministry of Science, 
Research and Arts and the Universities of the State of Baden-Württemberg, Germany, within the framework program bwHPC.

\bibliography{CRM_REVTEX}

\begin{thebibliography}{21}%
\makeatletter
\providecommand \@ifxundefined [1]{%
 \@ifx{#1\undefined}
}%
\providecommand \@ifnum [1]{%
 \ifnum #1\expandafter \@firstoftwo
 \else \expandafter \@secondoftwo
 \fi
}%
\providecommand \@ifx [1]{%
 \ifx #1\expandafter \@firstoftwo
 \else \expandafter \@secondoftwo
 \fi
}%
\providecommand \natexlab [1]{#1}%
\providecommand \enquote  [1]{``#1''}%
\providecommand \bibnamefont  [1]{#1}%
\providecommand \bibfnamefont [1]{#1}%
\providecommand \citenamefont [1]{#1}%
\providecommand \href@noop [0]{\@secondoftwo}%
\providecommand \href [0]{\begingroup \@sanitize@url \@href}%
\providecommand \@href[1]{\@@startlink{#1}\@@href}%
\providecommand \@@href[1]{\endgroup#1\@@endlink}%
\providecommand \@sanitize@url [0]{\catcode `\\12\catcode `\$12\catcode
  `\&12\catcode `\#12\catcode `\^12\catcode `\_12\catcode `\%12\relax}%
\providecommand \@@startlink[1]{}%
\providecommand \@@endlink[0]{}%
\providecommand \url  [0]{\begingroup\@sanitize@url \@url }%
\providecommand \@url [1]{\endgroup\@href {#1}{\urlprefix }}%
\providecommand \urlprefix  [0]{URL }%
\providecommand \Eprint [0]{\href }%
\providecommand \doibase [0]{http://dx.doi.org/}%
\providecommand \selectlanguage [0]{\@gobble}%
\providecommand \bibinfo  [0]{\@secondoftwo}%
\providecommand \bibfield  [0]{\@secondoftwo}%
\providecommand \translation [1]{[#1]}%
\providecommand \BibitemOpen [0]{}%
\providecommand \bibitemStop [0]{}%
\providecommand \bibitemNoStop [0]{.\EOS\space}%
\providecommand \EOS [0]{\spacefactor3000\relax}%
\providecommand \BibitemShut  [1]{\csname bibitem#1\endcsname}%
\let\auto@bib@innerbib\@empty
\bibitem [{\citenamefont {Aviram}\ and\ \citenamefont {Ratner}(1974)}]{aviram}%
  \BibitemOpen
  \bibfield  {author} {\bibinfo {author} {\bibfnamefont {A.}~\bibnamefont
  {Aviram}}\ and\ \bibinfo {author} {\bibfnamefont {M.~A.}\ \bibnamefont
  {Ratner}},\ }\href@noop {} {\bibfield  {journal} {\bibinfo  {journal}
  {Chem.~Phys.~Lett.}\ }\textbf {\bibinfo {volume} {29}} (\bibinfo {year}
  {1974})}\BibitemShut {NoStop}%
\bibitem [{\citenamefont {Geddes}\ \emph {et~al.}(1990)\citenamefont {Geddes},
  \citenamefont {Sandman}, \citenamefont {Sambles}, \citenamefont {Jarvis},\
  and\ \citenamefont {Parker}}]{exreali1}%
  \BibitemOpen
  \bibfield  {author} {\bibinfo {author} {\bibfnamefont {N.~J.}\ \bibnamefont
  {Geddes}}, \bibinfo {author} {\bibfnamefont {D.~J.}\ \bibnamefont {Sandman}},
  \bibinfo {author} {\bibfnamefont {J.~R.}\ \bibnamefont {Sambles}}, \bibinfo
  {author} {\bibfnamefont {D.~J.}\ \bibnamefont {Jarvis}}, \ and\ \bibinfo
  {author} {\bibfnamefont {W.~G.}\ \bibnamefont {Parker}},\ }\href@noop {}
  {\bibfield  {journal} {\bibinfo  {journal} {Appl.~Phys.~Lett.}\ }\textbf
  {\bibinfo {volume} {56}} (\bibinfo {year} {1990})}\BibitemShut {NoStop}%
\bibitem [{\citenamefont {Joachim}\ \emph {et~al.}(2000)\citenamefont
  {Joachim}, \citenamefont {Gimzewski},\ and\ \citenamefont
  {Aviram}}]{exreali3}%
  \BibitemOpen
  \bibfield  {author} {\bibinfo {author} {\bibfnamefont {C.}~\bibnamefont
  {Joachim}}, \bibinfo {author} {\bibfnamefont {J.~K.}\ \bibnamefont
  {Gimzewski}}, \ and\ \bibinfo {author} {\bibfnamefont {A.}~\bibnamefont
  {Aviram}},\ }\href@noop {} {\bibfield  {journal} {\bibinfo  {journal}
  {Nature}\ }\textbf {\bibinfo {volume} {408}} (\bibinfo {year}
  {2000})}\BibitemShut {NoStop}%
\bibitem [{\citenamefont {Zhao}\ \emph {et~al.}(2005)\citenamefont {Zhao},
  \citenamefont {Zeng}, \citenamefont {Cheng}, \citenamefont {Wang},
  \citenamefont {Wang}, \citenamefont {Yang}, \citenamefont {Hou},\ and\
  \citenamefont {Zhu}}]{exreali4}%
  \BibitemOpen
  \bibfield  {author} {\bibinfo {author} {\bibfnamefont {J.}~\bibnamefont
  {Zhao}}, \bibinfo {author} {\bibfnamefont {C.}~\bibnamefont {Zeng}}, \bibinfo
  {author} {\bibfnamefont {X.}~\bibnamefont {Cheng}}, \bibinfo {author}
  {\bibfnamefont {K.}~\bibnamefont {Wang}}, \bibinfo {author} {\bibfnamefont
  {G.}~\bibnamefont {Wang}}, \bibinfo {author} {\bibfnamefont {J.}~\bibnamefont
  {Yang}}, \bibinfo {author} {\bibfnamefont {J.}~\bibnamefont {Hou}}, \ and\
  \bibinfo {author} {\bibfnamefont {Q.}~\bibnamefont {Zhu}},\ }\href@noop {}
  {\bibfield  {journal} {\bibinfo  {journal} {Phys.~Rev.~Lett.}\ }\textbf
  {\bibinfo {volume} {95}} (\bibinfo {year} {2005})}\BibitemShut {NoStop}%
\bibitem [{\citenamefont {Martin}\ \emph {et~al.}(1993)\citenamefont {Martin},
  \citenamefont {Sambles},\ and\ \citenamefont {Ashwell}}]{exreali5}%
  \BibitemOpen
  \bibfield  {author} {\bibinfo {author} {\bibfnamefont {A.}~\bibnamefont
  {Martin}}, \bibinfo {author} {\bibfnamefont {J.}~\bibnamefont {Sambles}}, \
  and\ \bibinfo {author} {\bibfnamefont {G.}~\bibnamefont {Ashwell}},\
  }\href@noop {} {\bibfield  {journal} {\bibinfo  {journal} {Phys.~Rev.~Lett.}\
  }\textbf {\bibinfo {volume} {70}},\ \bibinfo {pages} {218} (\bibinfo {year}
  {1993})}\BibitemShut {NoStop}%
\bibitem [{\citenamefont {Metzger}\ \emph {et~al.}()\citenamefont {Metzger},
  \citenamefont {Chen}, \citenamefont {Hoepfner}, \citenamefont
  {Lakshmikantham}, \citenamefont {Vuillaume}, \citenamefont {Kawai},
  \citenamefont {Wu}, \citenamefont {Tachibana}, \citenamefont {Hughes},
  \citenamefont {Sakurai}, \citenamefont {Baldwin}, \citenamefont {Hosch},
  \citenamefont {Cava}, \citenamefont {Brehmer},\ and\ \citenamefont
  {Ashwell}}]{exreali6}%
  \BibitemOpen
  \bibfield  {author} {\bibinfo {author} {\bibfnamefont {R.}~\bibnamefont
  {Metzger}}, \bibinfo {author} {\bibfnamefont {B.}~\bibnamefont {Chen}},
  \bibinfo {author} {\bibfnamefont {U.}~\bibnamefont {Hoepfner}}, \bibinfo
  {author} {\bibfnamefont {M.}~\bibnamefont {Lakshmikantham}}, \bibinfo
  {author} {\bibfnamefont {D.}~\bibnamefont {Vuillaume}}, \bibinfo {author}
  {\bibfnamefont {T.}~\bibnamefont {Kawai}}, \bibinfo {author} {\bibfnamefont
  {X.}~\bibnamefont {Wu}}, \bibinfo {author} {\bibfnamefont {H.}~\bibnamefont
  {Tachibana}}, \bibinfo {author} {\bibfnamefont {T.}~\bibnamefont {Hughes}},
  \bibinfo {author} {\bibfnamefont {H.}~\bibnamefont {Sakurai}}, \bibinfo
  {author} {\bibfnamefont {J.}~\bibnamefont {Baldwin}}, \bibinfo {author}
  {\bibfnamefont {C.}~\bibnamefont {Hosch}}, \bibinfo {author} {\bibfnamefont
  {M.}~\bibnamefont {Cava}}, \bibinfo {author} {\bibfnamefont {L.}~\bibnamefont
  {Brehmer}}, \ and\ \bibinfo {author} {\bibfnamefont {G.}~\bibnamefont
  {Ashwell}},\ }\href@noop {} {\bibfield  {journal} {\bibinfo  {journal}
  {J.~Am.~Chem.~Soc.}\ }\textbf {\bibinfo {volume} {119}}}\BibitemShut
  {NoStop}%
\bibitem [{\citenamefont {Zhou}\ \emph {et~al.}()\citenamefont {Zhou},
  \citenamefont {Deshpande}, \citenamefont {Reed}, \citenamefont {Jones~II},\
  and\ \citenamefont {Tour}}]{exreali7}%
  \BibitemOpen
  \bibfield  {author} {\bibinfo {author} {\bibfnamefont {C.}~\bibnamefont
  {Zhou}}, \bibinfo {author} {\bibfnamefont {M.}~\bibnamefont {Deshpande}},
  \bibinfo {author} {\bibfnamefont {M.}~\bibnamefont {Reed}}, \bibinfo {author}
  {\bibfnamefont {L.}~\bibnamefont {Jones~II}}, \ and\ \bibinfo {author}
  {\bibfnamefont {J.}~\bibnamefont {Tour}},\ }\href@noop {} {\bibfield
  {journal} {\bibinfo  {journal} {Appl.~Phys.~Lett.}\ }\textbf {\bibinfo
  {volume} {71}}}\BibitemShut {NoStop}%
\bibitem [{\citenamefont {Mujica}\ \emph {et~al.}(2002)\citenamefont {Mujica},
  \citenamefont {Ratner},\ and\ \citenamefont {Nitzan}}]{discussion1}%
  \BibitemOpen
  \bibfield  {author} {\bibinfo {author} {\bibfnamefont {V.}~\bibnamefont
  {Mujica}}, \bibinfo {author} {\bibfnamefont {M.~A.}\ \bibnamefont {Ratner}},
  \ and\ \bibinfo {author} {\bibfnamefont {A.}~\bibnamefont {Nitzan}},\
  }\href@noop {} {\bibfield  {journal} {\bibinfo  {journal} {Chem.~Phys.}\
  }\textbf {\bibinfo {volume} {281}} (\bibinfo {year} {2002})}\BibitemShut
  {NoStop}%
\bibitem [{\citenamefont {{Kornilovitch}}\ \emph {et~al.}(2002)\citenamefont
  {{Kornilovitch}}, \citenamefont {{Bratkovsky}},\ and\ \citenamefont {{Stanley
  Williams}}}]{2002PhRvB..66p5436K}%
  \BibitemOpen
  \bibfield  {author} {\bibinfo {author} {\bibfnamefont {P.~E.}\ \bibnamefont
  {{Kornilovitch}}}, \bibinfo {author} {\bibfnamefont {A.~M.}\ \bibnamefont
  {{Bratkovsky}}}, \ and\ \bibinfo {author} {\bibfnamefont {R.}~\bibnamefont
  {{Stanley Williams}}},\ }\href {\doibase 10.1103/PhysRevB.66.165436}
  {\bibfield  {journal} {\bibinfo  {journal} {Phys.~Rev.~B}\ }\textbf {\bibinfo
  {volume} {66}},\ \bibinfo {eid} {165436} (\bibinfo {year} {2002})},\ \Eprint
  {http://arxiv.org/abs/arXiv:cond-mat/0206002} {arXiv:cond-mat/0206002}
  \BibitemShut {NoStop}%
\bibitem [{\citenamefont {Stokbro}\ \emph {et~al.}(2003)\citenamefont
  {Stokbro}, \citenamefont {Taylor},\ and\ \citenamefont
  {Brandbyge}}]{doi:10.1021/ja028229x}%
  \BibitemOpen
  \bibfield  {author} {\bibinfo {author} {\bibfnamefont {K.}~\bibnamefont
  {Stokbro}}, \bibinfo {author} {\bibfnamefont {J.}~\bibnamefont {Taylor}}, \
  and\ \bibinfo {author} {\bibfnamefont {M.}~\bibnamefont {Brandbyge}},\ }\href
  {\doibase 10.1021/ja028229x} {\bibfield  {journal} {\bibinfo  {journal}
  {Journal of the American Chemical Society}\ }\textbf {\bibinfo {volume}
  {125}},\ \bibinfo {pages} {3674} (\bibinfo {year} {2003})},\ \bibinfo {note}
  {pMID: 12656578}\BibitemShut {NoStop}%
\bibitem [{\citenamefont {Datta}\ \emph {et~al.}(1997)\citenamefont {Datta},
  \citenamefont {Tian}, \citenamefont {Hong}, \citenamefont {Reifenberger},
  \citenamefont {Henderson},\ and\ \citenamefont {Kubiak}}]{Datta}%
  \BibitemOpen
  \bibfield  {author} {\bibinfo {author} {\bibfnamefont {S.}~\bibnamefont
  {Datta}}, \bibinfo {author} {\bibfnamefont {W.}~\bibnamefont {Tian}},
  \bibinfo {author} {\bibfnamefont {S.}~\bibnamefont {Hong}}, \bibinfo {author}
  {\bibfnamefont {R.}~\bibnamefont {Reifenberger}}, \bibinfo {author}
  {\bibfnamefont {J.~I.}\ \bibnamefont {Henderson}}, \ and\ \bibinfo {author}
  {\bibfnamefont {C.~P.}\ \bibnamefont {Kubiak}},\ }\href {\doibase
  10.1103/PhysRevLett.79.2530} {\bibfield  {journal} {\bibinfo  {journal}
  {Phys. Rev. Lett.}\ }\textbf {\bibinfo {volume} {79}},\ \bibinfo {pages}
  {2530} (\bibinfo {year} {1997})}\BibitemShut {NoStop}%
\bibitem [{\citenamefont {Roy}(2010)}]{roy}%
  \BibitemOpen
  \bibfield  {author} {\bibinfo {author} {\bibfnamefont {D.}~\bibnamefont
  {Roy}},\ }\href {\doibase 10.1103/PhysRevB.81.085330} {\bibfield  {journal}
  {\bibinfo  {journal} {Phys. Rev. B}\ }\textbf {\bibinfo {volume} {81}},\
  \bibinfo {pages} {085330} (\bibinfo {year} {2010})}\BibitemShut {NoStop}%
\bibitem [{\citenamefont {Bransch{\"a}del}\ \emph {et~al.}(2010)\citenamefont
  {Bransch{\"a}del}, \citenamefont {Schneider},\ and\ \citenamefont
  {Schmitteckert}}]{ANDP:ANDP201000017}%
  \BibitemOpen
  \bibfield  {author} {\bibinfo {author} {\bibfnamefont {A.}~\bibnamefont
  {Bransch{\"a}del}}, \bibinfo {author} {\bibfnamefont {G.}~\bibnamefont
  {Schneider}}, \ and\ \bibinfo {author} {\bibfnamefont {P.}~\bibnamefont
  {Schmitteckert}},\ }\href {\doibase 10.1002/andp.201000017} {\bibfield
  {journal} {\bibinfo  {journal} {Annalen der Physik}\ }\textbf {\bibinfo
  {volume} {522}},\ \bibinfo {pages} {657} (\bibinfo {year}
  {2010})}\BibitemShut {NoStop}%
\bibitem [{\citenamefont {White}(1992)}]{PhysRevLett.69.2863}%
  \BibitemOpen
  \bibfield  {author} {\bibinfo {author} {\bibfnamefont {S.~R.}\ \bibnamefont
  {White}},\ }\href {\doibase 10.1103/PhysRevLett.69.2863} {\bibfield
  {journal} {\bibinfo  {journal} {Phys. Rev. Lett.}\ }\textbf {\bibinfo
  {volume} {69}},\ \bibinfo {pages} {2863} (\bibinfo {year}
  {1992})}\BibitemShut {NoStop}%
\bibitem [{\citenamefont {Schmitteckert}(2004)}]{PhysRevB.70.121302}%
  \BibitemOpen
  \bibfield  {author} {\bibinfo {author} {\bibfnamefont {P.}~\bibnamefont
  {Schmitteckert}},\ }\href {\doibase 10.1103/PhysRevB.70.121302} {\bibfield
  {journal} {\bibinfo  {journal} {Phys. Rev. B}\ }\textbf {\bibinfo {volume}
  {70}},\ \bibinfo {pages} {121302} (\bibinfo {year} {2004})}\BibitemShut
  {NoStop}%
\bibitem [{\citenamefont {Boulat}\ \emph {et~al.}(2008)\citenamefont {Boulat},
  \citenamefont {Saleur},\ and\ \citenamefont
  {Schmitteckert}}]{PhysRevLett.101.140601}%
  \BibitemOpen
  \bibfield  {author} {\bibinfo {author} {\bibfnamefont {E.}~\bibnamefont
  {Boulat}}, \bibinfo {author} {\bibfnamefont {H.}~\bibnamefont {Saleur}}, \
  and\ \bibinfo {author} {\bibfnamefont {P.}~\bibnamefont {Schmitteckert}},\
  }\href {\doibase 10.1103/PhysRevLett.101.140601} {\bibfield  {journal}
  {\bibinfo  {journal} {Phys. Rev. Lett.}\ }\textbf {\bibinfo {volume} {101}},\
  \bibinfo {pages} {140601} (\bibinfo {year} {2008})}\BibitemShut {NoStop}%
\bibitem [{\citenamefont {Datta}(2004)}]{0957-4484-15-7-051}%
  \BibitemOpen
  \bibfield  {author} {\bibinfo {author} {\bibfnamefont {S.}~\bibnamefont
  {Datta}},\ }\href {http://stacks.iop.org/0957-4484/15/i=7/a=051} {\bibfield
  {journal} {\bibinfo  {journal} {Nanotechnology}\ }\textbf {\bibinfo {volume}
  {15}},\ \bibinfo {pages} {S433} (\bibinfo {year} {2004})}\BibitemShut
  {NoStop}%
\bibitem [{\citenamefont {Lippmann}\ and\ \citenamefont
  {Schwinger}(1950)}]{Lippmann_Schwinger:PR50}%
  \BibitemOpen
  \bibfield  {author} {\bibinfo {author} {\bibfnamefont {B.~A.}\ \bibnamefont
  {Lippmann}}\ and\ \bibinfo {author} {\bibfnamefont {J.}~\bibnamefont
  {Schwinger}},\ }\href {\doibase 10.1103/PhysRev.79.469} {\bibfield  {journal}
  {\bibinfo  {journal} {Phys. Rev.}\ }\textbf {\bibinfo {volume} {79}},\
  \bibinfo {pages} {469} (\bibinfo {year} {1950})}\BibitemShut {NoStop}%
\bibitem [{\citenamefont {Landauer}(1996)}]{Landauer1}%
  \BibitemOpen
  \bibfield  {author} {\bibinfo {author} {\bibfnamefont {R.}~\bibnamefont
  {Landauer}},\ }\href {\doibase 10.1063/1.531590} {\bibfield  {journal}
  {\bibinfo  {journal} {Journal of Mathematical Physics}\ }\textbf {\bibinfo
  {volume} {37}},\ \bibinfo {pages} {5259–5268} (\bibinfo {year}
  {1996})}\BibitemShut {NoStop}%
\bibitem [{\citenamefont {Landauer}(1970)}]{doi:10.1080/14786437008238472}%
  \BibitemOpen
  \bibfield  {author} {\bibinfo {author} {\bibfnamefont {R.}~\bibnamefont
  {Landauer}},\ }\href {\doibase 10.1080/14786437008238472} {\bibfield
  {journal} {\bibinfo  {journal} {Philosophical Magazine}\ }\textbf {\bibinfo
  {volume} {21}},\ \bibinfo {pages} {863} (\bibinfo {year} {1970})},\ \Eprint
  {http://arxiv.org/abs/http://www.tandfonline.com/doi/pdf/10.1080/14786437008238472}
  {http://www.tandfonline.com/doi/pdf/10.1080/14786437008238472} \BibitemShut
  {NoStop}%
\bibitem [{\citenamefont {B\"uttiker}(1986)}]{PhysRevLett.57.1761}%
  \BibitemOpen
  \bibfield  {author} {\bibinfo {author} {\bibfnamefont {M.}~\bibnamefont
  {B\"uttiker}},\ }\href {\doibase 10.1103/PhysRevLett.57.1761} {\bibfield
  {journal} {\bibinfo  {journal} {Phys. Rev. Lett.}\ }\textbf {\bibinfo
  {volume} {57}},\ \bibinfo {pages} {1761} (\bibinfo {year}
  {1986})}\BibitemShut {NoStop}%
\end{thebibliography}%
\end{document}